\documentclass[prb,showpacs,amsmath,amssymb]{revtex4}
\usepackage{graphicx}

\usepackage{dcolumn}
\usepackage{bm}
\makeatother

\begin{document}

\title{A General Approach to Bosonization}
\author{Girish S. Setlur and V. Meera}
\affiliation{Department of Physics, Indian Institute of
Technology, Guwahati, \\North Guwahati, Assam 781 039, India}

\date{\today}

\begin{abstract}
We summarize recent developments in the field of higher
dimensional bosonization made by the authors and collaborators and
propose a general formula for the field operator in terms of
currents and densities in one dimension using a new ingredient
known as a `singular complex number'.
 Using this formalism, we compute the Green function of the
homogeneous electron gas in one spatial dimension with short-range
interaction leading to the Luttinger liquid and  also with
long-range interactions that leads to a Wigner crystal
 whose momentum distribution computed recently exhibits essential singularities.
 We generalize the formalism to
  finite temperature by combining with the author's hydrodynamic approach.
  The one-particle Green function of this system with essential singularities
   cannot be easily computed using the traditional approach to
 bosonization which involves the introduction of momentum cutoffs, hence the more general
  approach of the present formalism is proposed as a suitable alternative.
\end{abstract}



\maketitle

\section{Introduction}

   As the term  suggests, `bosonization' is an effort to recast theories involving entities
   that are not bosons in terms of bosons that are typically expressed in terms of bilinears of
   the original fields. Indeed it is even possible to recast theories involving bosons in terms of
    other bosons. This activity is not merely a pedantic exercise, for the end result of these efforts
   is a nonperturbative technique for studying the original theory. In particular, applying this technique
    to fermions in one spatial dimension leads to what are known as non (Landau)-Fermi liquids
     where the momentum distribution is continuous at the Fermi momentum.
   Thus one can bosonize fermions, spins
    and even complex scalar fields which are themselves bosons. Bosonization, according to our
     understanding, is nothing but the polar representation of a complex number.
      Bosonization of spins is accomplished by polar decomposing
     the ladder operators which leads to a semiclassical theory of spins.
    Complex scalar fields may be easily bosonized by polar decomposition as well. One may suspect that
     a similar decomposition should be feasible for fermions too. This has proved harder than one
      might hope. In one spatial dimension, the Thirring model which describes relativistic fermions
       self-interacting via a short-range repulsion was shown to be equivalent to the so-called Sine-Gordon
        theory which involves scalar fields\cite{Coleman}.
         This technique has been adapted to study condensed matter problems
         where the parabolic dispersion of the free fermions in Galilean invariant systems is linearized
          near the Fermi points so that the fermions now have linear dispersion and are moving with the Fermi
           velocity. The Fermi velocity takes on the role of the speed of light making the analogy complete.
 In more than one dimension, this program has not been particularly successful. In hindsight, it appears that it is
  premature to regard the framework available even in one spatial dimension as a closed subject.
   The Luttinger model caricature it seems, is unable to handle some exotic situations involving truly long range interactions
    in condensed matter systems. This was pointed out by the
    present author and this work is a summary and continuation of
    earlier works\cite{Setlur1}\cite{Setlur2}\cite{Setlur3}\cite{Setlur4}\cite{Setlur5}.
    The work of Schulz\cite{HJSchulz}on electrons interacting with
    potential ($ V(x) \sim 1/|x| $) using standard bosonization
    techniques is able to study the structure factor the Wigner crystal quite
    effectively but when it comes to the one-particle Green function, the
    results are very sketchy and cutoff dependent. It is not clear
    whether the method of Schulz can be used to study the one-particle
    Green function of a 1D Fermi system with potential $ V(x) \sim
    -|x| $, since our results show that the momentum distribution
    does not depend on arbitrary cutoffs, whereas they are
    mandatory in such standard bosonization methods.
    Furthermore, there are some technical subtleties involving so-called Klein
     factors that ensure fermion commutation rules between Fermi fields
      in the traditional approach that are far from satisfactory.
      The purpose of the present article is to
        summarize recent developments in the subject made by the author and his collaborators
 and to present a formula for the field operator in terms of currents and densities that is valid in a general
  sense. We stress that this write up is by no means a review of the entire field of bosonization,
   it only highlights the author's contributions, while not completely ignoring the work of others.
    We then go on to use this formula to calculate the Green functions of interacting
   systems in one spatial dimension, one with short range repulsion leading to the Luttinger liquid and also
    with a specific long range interaction which leads to the Wigner crystal.

     At this stage it is appropriate to
     survey some relevant literature on the subject of bosonization of fermions in general and higher dimensional
      bosonization in particular. This subject, as is well known, started with the papers of
       Tomonaga and Luttinger.
       Later on, Lieb and Mattis, Luther, Luther and Peschel developed it
       further. All these well-known literature may be found in the
       reviews and texts \cite{review1,review2}.
        This program was taken up by
       Haldane \cite{Haldane1} who coined the term `Luttinger liquid' to describe systems whose Green functions
        have power law singularities rather than simple poles. The generalization of these ideas to
         higher dimensions was attempted by Luther. The effort to generalize these ideas to higher
          dimensions seemed phenomenologically ill-founded as one does not expect to see
           non-Fermi liquids in higher dimensions except perhaps for very long-range interactions
            which are un-physical. However the phenomenon of high temperature superconductivity suggested the need
             for such a ground state. In the mid-nineties starting from the work of
              Haldane \cite{Haldane2}, Castro-Neto and Fradkin\cite{Castro},
              Houghton et. al.\cite{Houghton}, Kopietz and collaborators\cite{Kopietz} attempted to develop a theory
              in higher dimensions along the lines of the Tomonaga-Luttinger theory.
               The works of Kopietz and collaborators
               deserve special mention since they have persisted with this technique over the years.
                These efforts closely mimic
               the theory in one dimension that is valid only in the sense of the random phase approximation (RPA)
               \cite{footnote2}.
                Attempts to go beyond this approximation using the original formalism seems futile. Hence Setlur
                 and collaborators developed a new scheme loosely based on the work of Castro-Neto and
                  Fradkin\cite{Castro} to overcome these difficulties. In what follows, we describe the outcome of
                   these efforts made by the present author and his collaborators and go on to write
                    down a formula for the field operator in terms of currents and densities.

\section{Review of the General Formalism}

In what follows we describe briefly the main results in the
authors' earlier works. It must be stressed that this write up is
not a substitute for a reading of those works. Let $ c_{ {\bf{k}}
} $ and $ c^{\dagger}_{ {\bf{k}} } $ be fermion annihilation and
creation operators. We may define new operators using these that
are called sea-bosons. They are in general, complicated non-local
combination of number conserving products of Fermi
fields\cite{Setlur2}. However in the sense of the random phase
approximation(RPA) we may write,
\begin{equation}
A_{ {\bf{k}} }({\bf{q}}) \approx
n_{F}({\bf{k}}-{\bf{q}}/2)(1-n_{F}({\bf{k}}+{\bf{q}}/2))\mbox{
}c^{\dagger}_{ {\bf{k}}-{\bf{q}}/2 }c_{ {\bf{k}}+{\bf{q}}/2 }
\end{equation}
Here $ n_{F}({\bf{p}}) = \theta(k_{F}-|{\bf{p}}|) $ is the momentum
distribution of free fermions. This object $ A_{ {\bf{k}}
}({\bf{q}}) $ has been shown to obey the following commutation
rules\cite{Setlur2}.
\begin{equation}
[A_{ {\bf{k}} }({\bf{q}}),A^{\dagger}_{ {\bf{k}}^{'} }({\bf{q}}^{'})] = n_{F}({\bf{k}}-{\bf{q}}/2)(1-n_{F}({\bf{k}}+{\bf{q}}/2))
 \mbox{    }\delta_{ {\bf{k}}, {\bf{k}}^{'} } \delta_{ {\bf{q}}, {\bf{q}}^{'} } \mbox{             }; \mbox{          }
[A_{ {\bf{k}} }({\bf{q}}),A_{ {\bf{k}}^{'} }({\bf{q}}^{'})] =
0\mbox{             }; \mbox{          } [A^{\dagger}_{ {\bf{k}}
}({\bf{q}}),A^{\dagger}_{ {\bf{k}}^{'} }({\bf{q}}^{'})] =
0\label{COMM}
\end{equation}
The defining equation for $ A_{ {\bf{k}} }({\bf{q}}) $ may be
partially inverted and a formula for the number conserving product
of two Fermi fields may be written down in terms of these
sea-bosons. Define $ c_{ {\bf{k}}, < } = n_{F}({\bf{k}}) \mbox{
}c_{ {\bf{k}} } $ and $ c_{ {\bf{k}}, > } = (1-n_{F}({\bf{k}}))
\mbox{ }c_{ {\bf{k}} } $. At the level of RPA we may write, $
c^{\dagger}_{ {\bf{k}} - {\bf{q}}/2, < }c_{ {\bf{k}}+{\bf{q}}/2, >
} \approx A_{ {\bf{k}} }({\bf{q}}) $ and $ c^{\dagger}_{ {\bf{k}}
+ {\bf{q}}/2, < } c_{ {\bf{k}}-{\bf{q}}/2, < } \approx 0 $ and
 $ c^{\dagger}_{ {\bf{k}}
+ {\bf{q}}/2, > } c_{ {\bf{k}}-{\bf{q}}/2, > } \approx 0 $.
 Thus at the RPA level $ c^{\dagger}_{ {\bf{k}}+{\bf{q}}/2 }c_{ {\bf{k}}-{\bf{q}}/2
 } \approx
 A_{ {\bf{k}} }(-{\bf{q}})
 + A^{\dagger}_{ {\bf{k}} }({\bf{q}}) $.
 In general (beyond RPA) we may write\cite{Setlur2} ($ {\bf{q}} \neq 0
$),
\begin{equation}
c^{\dagger}_{ {\bf{k}}+{\bf{q}}/2 }c_{ {\bf{k}}-{\bf{q}}/2 }\approx
 A_{ {\bf{k}} }(-{\bf{q}})
 + A^{\dagger}_{ {\bf{k}} }({\bf{q}})
  + \sum_{ {\bf{q}}_{1} } A^{\dagger}_{ {\bf{k}} + {\bf{q}}/2 - {\bf{q}}_{1}/2 }({\bf{q}}_{1})
  A_{ {\bf{k}} - {\bf{q}}_{1}/2 }(-{\bf{q}}+{\bf{q}}_{1})
   -  \sum_{ {\bf{q}}_{1} } A^{\dagger}_{ {\bf{k}} - {\bf{q}}/2 + {\bf{q}}_{1}/2 }({\bf{q}}_{1})
  A_{ {\bf{k}} + {\bf{q}}_{1}/2 }(-{\bf{q}}+{\bf{q}}_{1})
\label{INV}
\end{equation}
 The above identification together with Eq.(\ref{COMM}) can be
 shown to be sufficient to reproduce the exact commutators between
 Fermi bilinears. The correspondence in Eq.(\ref{INV}) is also
 sufficient to reproduce all the dynamical correlation functions of the operator
  $ c^{\dagger}_{ {\bf{k}}+{\bf{q}}/2 }c_{ {\bf{k}}-{\bf{q}}/2 } $ of
 the free theory provided we set the kinetic energy operator to be
 $ K = \sum_{ {\bf{k}} {\bf{q}} } \left( \frac{ {\bf{k.q}} }{m}
 \right)A^{\dagger}_{ {\bf{k}} }({\bf{q}})A_{ {\bf{k}} }({\bf{q}})
 $ [for more details please consult our published works].
 Thus any theory involving fermions that conserves their total
number may be reexpressed in terms of these bosons. The main
purpose of the present article is to {\it{fully}} invert
 the defining equation for $ A_{ {\bf{k}} }({\bf{q}}) $ and express the field operator $ c_{ {\bf{p}} } $ alone in terms
  of these bosons. To accomplish this, we first attempt to polar decompose the field operator in real space
   $ \psi({\vec{r}}) = \frac{1}{ \sqrt{V} } \sum_{ {\bf{p}} } e^{i {\bf{p}} \cdot {\vec{r}} } c_{ {\bf{p}} } $.
\begin{equation}
\psi({\vec{r}}) = e^{ i \Lambda([\rho];{\vec{r}}) } e^{ -i \Pi({\vec{r}}) } \sqrt{ \rho({\vec{r}}) }
\label{DPVA}
\end{equation}
Here we have introduced a new variable $ \Pi({\vec{r}}) $ which is a
canonical conjugate to $ \rho({\vec{r}}) = \psi^{\dagger}({\vec{r}})
\psi({\vec{r}}) $\cite{footnote}. This means that $ [\Pi({\vec{r}}),
\Pi({\vec{r}}^{'})] = 0 $ and
 $ [\Pi({\vec{r}}), \rho({\vec{r}}^{'})] = i \delta({\vec{r}}-{\vec{r}}^{'}) $ and of course
  $ [ \rho({\vec{r}}), \rho({\vec{r}}^{'})] = 0 $. As an operator we know that $ \rho $ is non-negative.
   It is then well known that $ \Pi $ cannot be self-adjoint. In fact we may write
  $ \Pi({\vec{r}}) = X_{0} + {\tilde{\Pi}}({\vec{r}}) $ where $  {\tilde{\Pi}}({\vec{r}}) $ is strictly self-adjoint
   and $ X_{0} $ is conjugate to the total number $ [X_{0}, {\hat{N}}] = i $.
 In what follows
we regard the number operator to be equal to infinity
(thermodynamic limit) in which case $ X_{0}
 $ should be arbitrarily small in order for it to be a conjugate to the number operator.
  Hence we shall not be too careful and we shall treat $ X_{0} $ as a
  real c-number and ignore it altogether.

We now divert our attention and examine the property of the fermion current (density) operator
 $ {\bf{J}}({\vec{r}}) = Im[ \psi^{\dagger}({\vec{r}}) \nabla \psi({\vec{r}}) ] $.
 In two and three dimensions,
  we may construct the operator $ {\bf{W}} = \rho (\nabla \times {\bf{J}}) - \nabla \rho \times {\bf{J}} $.
  We first represent currents and densities in real space
   $ \rho({\vec{r}}) = \sum_{i} \delta({\vec{r}}-{\vec{r}}_{i}) $ and
    $ {\bf{J}}({\vec{r}}) = \sum_{i}\frac{ -i \nabla_{i} }{2} \delta({\vec{r}}-{\vec{r}}_{i})
    + \sum_{i}\delta({\vec{r}}-{\vec{r}}_{i}) \frac{ -i \nabla_{i} }{2}   $.
   By acting this on fermionic wavefunctions in real space we conclude that $ {\bf{W}} \equiv 0 $. This means
    that $ {\bf{W}} = \rho^2 \nabla \times \frac{1}{ \rho }{\bf{J}} = 0 $. In other words, there exists a scalar
     $ \Pi $ such that $ {\bf{J}} = - \rho \nabla \Pi $. Independently we may argue that
     a more general ansatz $ {\bf{J}} = - \rho \nabla \Pi + {\bf{C}}([\rho];{\vec{r}}) $ fails to reproduce the
      current-current commutator unless $ {\bf{C}} \equiv 0 $.
       It appears then, that the function $ \Lambda $ in Eq.(\ref{DPVA}) should be independent of $ {\vec{r}} $.
 This $ \Lambda $ is crucial since it determines the statistics of the field $ \psi $.
  In particular, setting
  $ \Lambda = 0 $ describes bosons rather than fermions.
  In our earlier work\cite{Setlur5}, we argued that the ansatz
   in Eq.(\ref{DPVA}) may be used  to derive an action in terms of $ \Pi $ and $ \rho $.
    In the lagrangian formalism,
    there are no operators. We demote the operators $ \Pi $ and $ \rho $ to the
     status of real numbers and
     use Eq.(\ref{DPVA}) in the action for free fermions,
      $ S = \int^{-i\beta}_{0}dt \int d^d x \mbox{        }\psi^{\dagger} (i \partial_{t} + \frac{ \nabla^2 }{2m}) \psi $.
      This led to the following action for free fermions,
 \begin{equation}
S =  \int^{-i\beta}_{0}dt \int d^d x  \left( \rho \partial_{t} \Pi -  V_{F}([\rho];{\bf{x}}) -
\frac{ \rho (\nabla \Pi)^2 + \frac{ ( \nabla \rho )^2 }{ 4 \rho }
}{ 2m } \right) \label{ACTION}
\end{equation}
 Here $ V_{F} $ is a functional of the density that has to be fixed
 by making contact with the properties of the
 free theory. We have shown that the RPA limit of the above action may be
 rigorously derived using sea-bosons\cite{Setlur5}.
 We have also shown \cite{Setlur5} that this leads to the
 following expression for the generating function of density correlations of a homogeneous
 electron gas in terms of the corresponding quantity for the free
 theory. If $ Z([U]) $ is the function that generates density
 correlations of the homogeneous electron gas where the electrons interact with a potential $ v_{ {\bf{q}} } $
  and $ Z_{0}([U]) $ is the corresponding quantity of the free theory ( for example $ \left( \frac{\delta^2 Z([U])}{ \delta U_{
 {\bf{q}}, n } \delta U_{ -{\bf{q}}, -n } } \right)_{ U \equiv 0 } \equiv <\rho_{
 {\bf{q}}, n } \rho_{ -{\bf{q}}, -n }> $ and $ <T \rho({\bf{q}},t)
 \rho(-{\bf{q}},t^{'})> = \sum_{n} e^{ - w_{n}(t-t^{'}) } <\rho_{
 {\bf{q}}, n } \rho_{ -{\bf{q}}, -n }> $  and $ w_{n} = 2 \pi n/
 \beta $ is the Matsubara frequency ) then,
\begin{equation}
Z([U]) = \int D[U^{'}] \mbox{    }e^{ \sum_{ {\bf{q}}n } \frac{ V
}{ 2 \beta v_{ {\bf{q}} } }(U_{ {\bf{q}} n } - U^{'}_{ {\bf{q}} n
})(U_{ -{\bf{q}}, -n } - U^{'}_{ -{\bf{q}}, -n }) }\mbox{
}Z_{0}([U^{'}])
\end{equation}
 It appears that we have to judiciously combine the hamiltonian or the operator version and the lagrangian
  version in order to obtain useful results. Thus we wish to now revert to the operator description to try and
   express the field operator explicitly in terms of currents and densities.
    We observed that current-current commutator implies that $ \Lambda $ is independent of $ {\vec{r}} $.
  Unfortunately, this conflicts with the requirement that $ \psi $ obey fermion commutation rules. Indeed, imposing these
  rules on $ \psi $ in Eq.(\ref{DPVA}) leads to the following constraint on $ \Lambda $.
\begin{equation}
e^{ i \Lambda([\rho];{\vec{r}}) }  e^{ i \Lambda([\{\rho({\vec{x}}) + \delta({\vec{x}}-{\vec{r}})\}];{\vec{r}}^{'}) }
  = - \mbox{   }e^{ i \Lambda([\rho];{\vec{r}}^{'}) }  e^{ i \Lambda([\{\rho({\vec{x}}) + \delta({\vec{x}}-{\vec{r}}^{'})
  \}];{\vec{r}}) }
\label{RECUR}
\end{equation}
If $ \Lambda $ is independent of $ {\vec{r}} $ this is impossible. Hence we seem to have reached an impasse.
There is a way out this difficulty  using the notion of what may be called `singular complex numbers'.
 We describe this concept in the following section.

\section{Field Operator Using Singular Complex Numbers}

  In our earlier work\cite{Setlur5} we argued that the field operator
  in momentum space may be expressed directly
  in terms of the sea-bosons provided we invoke the concept of a singular complex number :
   $ w_{ {\bf{p}} } = e^{-i N^{0}\xi_{ {\bf{p}} } } $, where $ N^{0} \rightarrow \infty $
    and $ \xi_{ {\bf{p}} } $ is arbitrary. Therefore,
$ w_{ {\bf{p}} }{\bar{w}}_{ {\bf{p}}^{'} } = {\bar{w}}_{
{\bf{p}}^{'} }w_{ {\bf{p}} } = \delta_{ {\bf{p}}, {\bf{p}}^{'} }
$. This rather unusual quantity may be motivated using the
following argument. Consider the number operator $ n_{ {\bf{k}} }
$. We may introduce formally a  conjugate $ P_{ {\bf{k}} } $
namely an operator that obeys $ [P_{ {\bf{k}} }, P_{ {\bf{k}}^{'}
}] = 0 $ and $ [P_{ {\bf{k}} }, n_{ {\bf{k}}^{'} }] = i \mbox{
       }\delta_{ {\bf{k}}, {\bf{k}}^{'} } $. If we are going to treat $ n_{ {\bf{k}} } $ as a c-number namely,
        $ n_{ {\bf{k}} } = n_{F}({\bf{k}}) \mbox{   } {\bf{1}} $, then we have to ensure that $ P_{ {\bf{k}} } $
         is  a formally infinite c-number in order that it is a conjugate to $ n_{ {\bf{k}} } $.
Thus $ w_{ {\bf{k}} } = e^{ -i P_{ {\bf{k}}} }  $ has the
properties that we have just described.
          We argue that the term $ e^{
i \Lambda } $ may be rewritten using these complex numbers. Let us
invoke the following
 ansatz that is inspired from our early work\cite{Setlur1} :
\begin{equation}
\psi({\vec{r}}) =  \left( \frac{1}{ \sqrt{N^{0}} }\sum_{ {\bf{p}} } e^{ i E_{ {\bf{p}} }([\rho]; {\vec{r}}) }\mbox{    }
 w_{ {\bf{p}} } \mbox{   }e^{i k_{F} {\hat{p}} \cdot {\vec{r}} } \mbox{   }n_{F}({\bf{p}}) \right)
\mbox{      }e^{ -i \Pi({\vec{r}}) }\mbox{   }
\sqrt{ \rho({\vec{r}}) }
\label{PSI}
\end{equation}
\begin{equation}
\psi^{\dagger}({\vec{r}}) =  \sqrt{ \rho({\vec{r}}) }\mbox{   }e^{ i \Pi({\vec{r}}) }
\mbox{   }\left( \frac{1}{ \sqrt{N^{0}} }\sum_{ {\bf{p}} } e^{ -i E_{ {\bf{p}} }([\rho]; {\vec{r}}) }\mbox{    }
 {\bar{w}}_{ {\bf{p}} } \mbox{   }e^{-i k_{F} {\hat{p}} \cdot {\vec{r}} } \mbox{   }n_{F}({\bf{p}}) \right)
\label{PSIDAG}
\end{equation}
 Here $ E_{ {\bf{p}} }([\rho]; {\vec{r}}) $ is real.
 The idea is that the rapidly varying part is written separately as a multiplying exponent
  $ e^{ \pm i k_{F} {\hat{p}} \cdot {\vec{r}} } $. The slowly varying portion is in the density and phase variables.
   Notice that the theory presented here is very general and there are no momentum cutoffs at the outset. Therefore
    the terms `slow' and `fast' are merely suggestive of approximations that will have to be used in the practical
     computations where such a distinction acquires concrete meaning.
If we postulate that $ \nabla E_{ {\bf{p}} }([\rho];{\vec{r}}) =
 - \nabla E_{ -{\bf{p}} }([\rho];{\vec{r}}) $
 then we find that,
$ \rho({\vec{r}}) = \psi^{\dagger}({\vec{r}}) \psi({\vec{r}})  $ and
$ {\bf{J}}({\vec{r}}) = Im[\psi^{\dagger}({\vec{r}}) (\nabla \psi({\vec{r}}))] = -\rho({\vec{r}}) \nabla \Pi({\vec{r}}) $.
Therefore, current algebra - the mutual commutation rules between currents and densities is then trivially obeyed.
 In other words,
$ e^{ i \Lambda([\rho];{\vec{r}}) } = \frac{1}{ \sqrt{N^{0}} }\sum_{ {\bf{p}} }
e^{ i E_{ {\bf{p}} }([\rho]; {\vec{r}}) }\mbox{    }
 w_{ {\bf{p}} } \mbox{   }e^{i k_{F}{\hat{p}} \cdot {\vec{r}} } \mbox{   }n_{F}({\bf{p}}) $
is unitary. Now we have to apply fermion commutation rules. We expect to derive a recursion relation similar
 to Eq.(\ref{RECUR}). Indeed we find that both the requirement $ \{\psi({\vec{r}}), \psi({\vec{r}}^{'})\} = 0 $
  and $  \{\psi({\vec{r}}), \psi^{\dagger}({\vec{r}}^{'})\} = \delta({\vec{r}}-{\vec{r}}^{'})  $
   are obeyed if we ensure that the following recursion holds,
\begin{equation}
 e^{ -i E_{ {\bf{p}} }([\{\rho({\vec{x}})+\delta({\vec{x}}-{\vec{r}}^{'})];\mbox{   }{\vec{r}}) }
  e^{ i E_{ {\bf{p}} }([\rho];\mbox{   }{\vec{r}}) }\mbox{    }
= -  e^{ -i E_{ {\bf{p}}^{'} }([\{\rho({\vec{x}})+\delta({\vec{r}}-{\vec{x}})\}];
\mbox{   }{\vec{r}}^{'}) }
\mbox{    }e^{ i E_{ {\bf{p}}^{'} }([\rho]; \mbox{   }{\vec{r}}^{'}) }
\end{equation}
In one dimension, we try the following ansatz. Note that $
{\hat{p}} = \pm 1 $ in one dimension. $  E_{ p^{'} }([\rho];
\mbox{ }r^{'})  = \int_{ -\infty }^{ \infty } dx \mbox{
}\rho(x)\mbox{ } D_{ {\hat{p}}^{'}}(x,r^{'}) $. The recursion
relation implies, $ e^{ -i  \mbox{  } D_{ {\hat{p}} }(r^{'},r)} =
-  e^{ -i  \mbox{  } D_{ {\hat{p}}^{'} }(r,r^{'})} $ whereas the
unitarity condition forces us to choose, $ D_{
{\hat{p}}^{'}}(x,r^{'}) = -{\hat{p}}^{'} D_{red}(x,r^{'}) $. Both
these are satisfied by the choice
 $ D_{ {\hat{p}}^{'}}(x,r^{'}) = -\pi {\hat{p}}^{'} \theta(x-r^{'}) $ where $ \theta(x) $ is the Heaviside
  step function ( $ \theta(x \leq 0) = 0 $ and $ \theta(x > 0) = 1  $ ).
   We have to verify that this choice reproduces the Green function of the noninteracting theory.
   This is shown later on. In higher dimensions we similarly try,
   $
E_{ {\bf{p}} }([\rho];\mbox{   }{\vec{r}}) =  \int d^dx \mbox{
}\rho({\vec{x}})\mbox{  } D_{ {\hat{p}} }({\vec{x}},{\vec{r}}) $.
 The recursion condition in higher dimensions : $ e^{ -i  \mbox{  }
D_{ {\hat{p}} }({\vec{r}}^{'},{\vec{r}})}
= -  e^{ -i  \mbox{  }
D_{ {\hat{p}}^{'} }({\vec{r}},{\vec{r}}^{'})} $
 seems rather hard to satisfy since $ {\hat{p}} $ and $ {\hat{p}}^{'} $ can
  be parallel to each other, antiparallel, or anything in between. If we are willing to
   require that only $ \{\psi({\vec{r}}), \psi^{\dagger}({\vec{r}}^{'}) \} = \delta({\vec{r}}-{\vec{r}}^{'}) $ be obeyed
    then a simple choice might be sufficient.
 In any event it is the propagator that we are interested in and for this it is this rule that is important.
 In this case $ {\hat{p}} = {\hat{p}}^{'} $ and we may choose, $
D_{ {\hat{p}} }({\vec{r}},{\vec{r}}^{'}) = \pi \mbox{ }\theta(
{\hat{p}} \cdot ({\vec{r}}-{\vec{r}}^{'}) ) $.
 This choice obeys both the unitarity condition :
  $ \nabla D_{ {\hat{p}} }({\vec{r}},{\vec{r}}^{'}) = -\nabla D_{ -{\hat{p}} }({\vec{r}},{\vec{r}}^{'}) $
   and the recursion.
The recursion breaks down when $ {\hat{p}} \cdot ({\vec{r}}-{\vec{r}}^{'}) = 0 $ but these are a set of points of measure
 zero. Unfortunately, it can be shown that this choice fails to reproduce the free propagator.
 This suggests that the function $ E_{ {\bf{p}} } $ may be a nonlinear function
    of the density in higher dimensions. Therefore in higher dimensions it is better to try and express
     the field variable directly in terms of sea-bosons.
     A partially correct formula has been provided in an earlier
     work\cite{Setlur5}, it appears that more work is needed to make that agenda
     practically useful. For now we use the expression for the field in terms
     of the hydrodynamic variables to calculate the Green
     function in one dimension, first of the Luttinger liquid. In the case of the Wigner crystal we find that a
     reinterpretation is needed that essentially involves re-expressing the field directly in terms of sea-bosons.
       For  example we may rewrite the field in one dimension as follows.
\begin{equation}
\psi(r,t) \approx  \left( \frac{e^{-i\epsilon_{F} t }}{ \sqrt{L}
}\sum_{ p } e^{\sum_{q} e^{ -i q r } [-\frac{ {\hat{p}}\pi}{ qL}
\rho_{q}(t)-i X_{-q}(t)] }\mbox{ }
 w_{ p } \mbox{   }e^{i k_{F} {\hat{p}}r } \mbox{   }n_{F}(p) \right)
\end{equation}
 The rapidly varying $ e^{-i\epsilon_{F} t } $ has to put in by
 hand, however it may be motivated by realizing that the $ X_{q=0}
 $ term is conjugate to the total number of particles and this
 picks up a contribution similar to the one suggested upon time
 evolution with respect to the hamiltonian of the system since, for example, the hamiltonian of the free Fermi theory
 may be written as  $ H = N\epsilon_{0} + \sum_{k,q} \frac{k.q}{m}A^{\dagger}_{k}(q)A_{k}(q) $.
  If we set $ \epsilon_{0} = \epsilon_{F} $ then we recover the
  factor suggested.
We write $ \rho_{q} = \sum_{k} [A_{k}(-q)+A^{\dagger}_{k}(q)] =
\rho_{R}(q) + \rho_{L}(q) $, where $ \rho_{R}(q) =\sum_{k>0}
[A_{k}(-q)+A^{\dagger}_{k}(q)] $ and
 $ \rho_{L}(q) =\sum_{k<0}
[A_{k}(-q)+A^{\dagger}_{k}(q)] $
 and $ X_{q} = \frac{ ik_{F} q }{ N q^2 }(\rho_{R}(-q) -
 \rho_{L}(-q))
 $. If $ {\hat{p}} = + 1 $ then $  [-\frac{ \pi}{ qL}  \rho_{q}-i
X_{-q}] = -\frac{2\pi}{ qL}\rho_{R}(q)  $. If $ {\hat{p}}= - 1 $
we have $ [\frac{ \pi}{ qL}  \rho_{q}-i X_{-q}] = \frac{2\pi}{ qL}
\rho_{L}(q) $. Thus we have the familiar result for right and left
movers. We may write $ \psi(r,t) = \psi_{R}(r,t) + \psi_{L}(r,t) $
or,
\begin{equation}
\psi(r) \approx  \left( \frac{1}{ \sqrt{L} }\sum_{ p>0 }
e^{-\sum_{q} e^{ -i q r } \frac{2\pi}{ qL}\rho_{R}(q) }\mbox{ }
 w_{ p } \mbox{   }e^{i k_{F} r } \mbox{   }n_{F}(p) \right)
 +\left( \frac{1}{ \sqrt{L} }\sum_{ p<0 }
e^{\sum_{q} e^{ -i q r } \frac{2\pi}{ qL} \rho_{L}(q) }\mbox{ }
 w_{ p } \mbox{   }e^{-i k_{F} r } \mbox{   }n_{F}(p) \right)
 \label{RNL}
\end{equation}
As usual we have $ [\rho_{R}(q),\rho_{R}(-q)] =
\sum_{k>0}[A_{k}(-q),A^{\dagger}_{k}(-q)] +
\sum_{k>0}[A^{\dagger}_{k}(q),A_{k}(q)] = - \frac{qL }{2\pi}$ and
 $ [\rho_{L}(q),\rho_{L}(-q)] = \frac{qL }{2\pi} $. This
 expression is well-known to the traditional bosonizing community
 and is known to reproduce the exponents of the Luttinger liquid
 correctly. In passing, we note that the present formalism does not
 allow terms such as $ \psi^{\dagger}_{R}(x) \psi_{L}(x^{'}) $,
 this being identically zero due to the singular complex number $
 w_{p} $. Thus in our formalism, the terms responsible for backward
 scattering come from the quadratic corrections on the right hand
 side of Eq.(\ref{INV}). Backward scattering is synonymous with
 large momentum transfer, which in turn means corrections to RPA,
 which then translates to quadratic corrections in Eq.(\ref{INV}).
  Now we wish to compute the Green function of the Wigner
  crystal whose momentum distribution has been computed in an earlier
  work\cite{Setlur4}. There we considered electrons on a circle interacting
   via a long range interaction $ V(x) = - \frac{ e^2 }{a^2}|x| $, where $ |x| $ is the chord length.
 The ground state of this system was shown to be crystalline with
 lattice spacing $ l_{c} = \pi/k_{F} $.
    It was shown that the momentum distribution of the Wigner crystal at zero temperature is given by,
\begin{equation}
{\bar{n}}_{p} = \frac{1}{2}\left( 1 + e^{ - \frac{ m \omega_{0}
}{k^2_{F}-p^2 } } \right)n_{F}(p)
 + \frac{1}{2}\left( 1 + e^{ - \frac{ m \omega_{0}
}{p^2-k_{F}^2 } } \right)(1-n_{F}(p)) \label{WIGNER}
\end{equation}
where, $ \omega_{0} = \sqrt{ \frac{ 2 e^2k_{F} }{ \pi m a^2 } } $.
This was derived using the general formula for the momentum
distribution in terms of sea-bosons,
\begin{equation}
{\bar{n}}_{ {\bf{p}} } = n_{F}({\bf{p}})\mbox{  }\frac{1}{2}
\left( 1 + e^{ - 2 \sum_{ {\bf{q}} }<A^{\dagger}_{
{\bf{p}}+{\bf{q}}/2 }({\bf{q}})A_{ {\bf{p}}+{\bf{q}}/2
}({\bf{q}})> } \right)
 + (1-n_{F}({\bf{p}}))\mbox{  }\frac{1}{2}
\left( 1 - e^{ - 2 \sum_{ {\bf{q}} }<A^{\dagger}_{
{\bf{p}}-{\bf{q}}/2 }({\bf{q}})A_{ {\bf{p}}-{\bf{q}}/2
}({\bf{q}})> } \right) \label{MOMDIS}
\end{equation}
First we wish to generalize Eq.(\ref{MOMDIS}) to finite
temperature. This is important since we know that even for a
noninteracting system, the momentum distribution at absolute zero
is discontinuous at $ k = k_{F} $ but is continuous at finite
temperature. A naive approach that just performs thermodynamic
averaging over the sea-boson occupation fails. We have to adopt a
more subtle approach. One possibility is to express the sea-bosons
in terms of the hydrodynamic variables and use the action in
Eq.(\ref{ACTION}). This can possibly be made to work out but is
really not worth the effort. It is better to gain some intuition
from this effort to guess the proper generalization. We propose
the following generalization.
\begin{equation}
{\bar{n}}_{ {\bf{p}} } = n_{F}({\bf{p}})\mbox{  }\frac{1}{2}
\left( 1 + \lambda({\bf{p}}) \mbox{  }e^{ - 2 \sum_{ {\bf{q}} }\ll
A^{\dagger}_{ {\bf{p}}+{\bf{q}}/2 }({\bf{q}})A_{
{\bf{p}}+{\bf{q}}/2 }({\bf{q}}) \gg } \right)
 + (1-n_{F}({\bf{p}}))\mbox{  }\frac{1}{2}
\left( 1 - \lambda({\bf{p}}) \mbox{  }e^{ - 2 \sum_{ {\bf{q}} }\ll
A^{\dagger}_{ {\bf{p}}-{\bf{q}}/2 }({\bf{q}})A_{
{\bf{p}}-{\bf{q}}/2 }({\bf{q}}) \gg } \right) \label{MOMGEN}
\end{equation}
Here $ \lambda({\bf{p}}) $ contains the temperature information of
the free theory only. The boson occupation in the exponent is
defined as follows : $ \ll ... \gg = < ... >_{\beta} - < ...
>_{\beta,0} $. That is, the difference between the interacting
theory at finite temperature and free theory at finite
temperature. To calculate the finite temperature sea-boson
occupation we have to invoke the hydrodynamic description. We
first express the sea-bosons in terms of hydrodynamic variables,
namely the velocity potential and density. It is given as follows
$ A_{ {\bf{k}} }({\bf{q}}) = [A_{ {\bf{k}} }({\bf{q}}),
A^{\dagger}_{ {\bf{k}} }({\bf{q}})] \left( \frac{ k_{F} }{ N |q| }
\rho_{ -{\bf{q}} } -i X_{ {\bf{q}} } \right) $. Here we have to
make sure that only the s-wave contributes, in other words we
ignore complications caused by the dot product $ {\bf{k.q}} $ and
replace it by its extremum value. This correspondence
automatically reproduces the RPA level identities also valid only
in the s-wave sense :
 $ \sum_{ {\bf{k}} }(A_{ {\bf{k}} }({\bf{q}}) + A^{\dagger}_{
 {\bf{k}} }(-{\bf{q}})) = \rho_{ -{\bf{q}} } $
  and $ \sum_{ {\bf{k}} }\frac{ i k_{F} }{ N |q| }
  (A_{ {\bf{k}} }({\bf{q}}) - A^{\dagger}_{
 {\bf{k}} }(-{\bf{q}})) = X_{ {\bf{q}} } $. We may also re-express
 these directly in terms of Fermi fields,
  $ \rho_{ {\bf{q}} }  = \sum_{ {\bf{k}} }c^{\dagger}_{ {\bf{k}}+{\bf{q}}/2 }c_{ {\bf{k}}-{\bf{q}}/2 } $
  and $ X_{ {\bf{q}} } = \sum_{ {\bf{k}} }\frac{ i k_{F} }{ N |q|
  }\mbox{     }sgn({\bf{k.q}})\mbox{    }
 c^{\dagger}_{ {\bf{k}}-{\bf{q}}/2 }c_{ {\bf{k}}+{\bf{q}}/2 }  $.
 This is beneficial since we may now easily compute the
 correlation functions of these operators at finite temperature. In general,
  $ <A^{\dagger}_{ {\bf{k}} }({\bf{q}})A_{ {\bf{k}} }({\bf{q}})> =
[A_{ {\bf{k}} }({\bf{q}}),A^{\dagger}_{ {\bf{k}} }({\bf{q}})]
\left( \frac{ \pi^2 }{ q^2L^2 } <\rho_{ {\bf{q}} } \rho_{
-{\bf{q}} }> + <X_{ -{\bf{q}} } X_{ {\bf{q}} }> - \frac{ \pi }{
|q|L } \right) $.  For a Luttinger liquid the zero temperature
theory is,
\begin{equation}
H = \sum_{k,q} \frac{k.q}{m}A^{\dagger}_{k}(q)A_{k}(q)
 + \sum_{q} \frac{ v_{0} }{L}
 [A(-q)A(q)+A^{\dagger}(q)A^{\dagger}(-q)] =
 \sum_{q}v|q|d^{\dagger}_{c}(q)d_{c}(q)
\end{equation}
where,
\begin{equation}
d_{c}(q) = \sqrt{ \frac{2\pi}{ |q|L }  }\left(\frac{v_{F}+v
}{2v}\right)^{\frac{1}{2}}A(q) + \sqrt{ \frac{2\pi}{ |q|L }
}\left(\frac{v_{F}-v }{2v}\right)^{\frac{1}{2}}A^{\dagger}(-q)
\end{equation}
\begin{equation}
d^{\dagger}_{c}(-q) = \sqrt{ \frac{2\pi}{ |q|L }
}\left(\frac{v_{F}+v }{2v}\right)^{\frac{1}{2}}A^{\dagger}(-q) +
\sqrt{ \frac{2\pi}{ |q|L } }\left(\frac{v_{F}-v
}{2v}\right)^{\frac{1}{2}}A(q)
\end{equation}
where $ v = \sqrt{ v_{F}^2 - \frac{ v^2_{0} }{ \pi^2 } } $ and $
[d_{c}(q),d^{\dagger}_{c}(q)] =  1 $ and $ \rho_{-q} = A(q) +
A^{\dagger}(-q) $ and $ X_{q} = \frac{ i \pi }{ |q|L
}(A(q)-A^{\dagger}(-q)) $ and $ A(q) = \sum_{k}A_{k}(q) $ and $
[A(q),A^{\dagger}(q)] =  \frac{ |q|L }{2\pi} $, $ \rho_{-q} =
(d_{c}(q) + d^{\dagger}_{c}(-q))c_{a} $
 and $ X_{q} =(d_{c}(q) - d^{\dagger}_{c}(-q))c_{b} $,
  $ c_{a} =  \sqrt{ \frac{ |q|L }{2\pi}  }\left( \left(\frac{v_{F}+v
}{2v}\right)^{\frac{1}{2}} - \left(\frac{v_{F}-v
}{2v}\right)^{\frac{1}{2}} \right) $ and $ c_{b} = \sqrt{ \frac{
|q|L }{2\pi}  }\left( \left(\frac{v_{F}+v
}{2v}\right)^{\frac{1}{2}} + \left(\frac{v_{F}-v
}{2v}\right)^{\frac{1}{2}} \right) \frac{ i \pi }{ |q|L } $. In
other words, $  <A^{\dagger}_{ k }(q)A_{ k }(q)> = [A_{ k
}(q),A^{\dagger}_{ k }(q)] ( \frac{v_{F}}{v} - 1 )\frac{ \pi }{
|q|L } $. Substituting this into Eq.(\ref{MOMDIS}) leads to a
momentum distribution that has power law singularities with
anomalous exponent $ \gamma = \frac{ v_{F} }{v} - 1 $. In general,
we have to use the RPA-level action in Eq.(\ref{ACTION}) to
calculate the finite temperature expectation values namely, $
<\rho_{ {\bf{q}} }\rho_{ -{\bf{q}} }> $ and $ <X_{ {\bf{q}} }X_{
-{\bf{q}} }> $. Let us now use this method to find the proper
generalization of Eq.(\ref{WIGNER}) to finite temperature. First,
we have to find $ \lambda({\bf{p}}) $. By making contact with the
free theory we find, $ \lambda({\bf{p}}) =
[sgn(k_{F}-|{\bf{p}}|)]^{-1}\mbox{ }Tanh \left[ \frac{ \beta
}{2}(\mu-\epsilon_{ {\bf{p}} }) \right] $. When this procedure is
implemented we obtain,
\begin{equation}
{\bar{n}}_{p} = \frac{1}{2}\left( 1 + \lambda(p) \mbox{   }e^{ -
\frac{ m \omega_{0} }{k^2_{F}-p^2 }coth \left( \frac{ \beta
\omega_{0} }{2}\right) } \right)n_{F}(p)
 + \frac{1}{2}\left( 1 -\lambda(p) \mbox{   } e^{ - \frac{ m \omega_{0}
}{p^2-k_{F}^2 }coth \left( \frac{ \beta \omega_{0} }{2}\right) }
\right)(1-n_{F}(p))
\end{equation}
Therefore the essential singularity remains even at finite
temperature. Finally we wish to calculate the dynamical propagator
of the Wigner crystal. It seems that in this case even
Eq.(\ref{PSI}) is not sufficient. We have to express the field
operator in momentum space directly in terms of the sea-bosons. A
partially correct formula was proposed in an earlier
work\cite{Setlur5}. Instead of searching for a rigorous approach,
we directly make the following surmise for the propagators that is
motivated by comparing with limiting cases.
 Define $ c_{ {\bf{p}}, < } = n_{F}({\bf{p}})\mbox{   }c_{
 {\bf{p}} } $ and $ c_{ {\bf{p}}, > } = (1-n_{F}({\bf{p}}))\mbox{   }c_{
 {\bf{p}} } $. Then,
\begin{equation}
<c^{\dagger}_{ {\bf{p}},> }(t^{'})c_{ {\bf{p}},> }(t)> =
(1-n_{F}({\bf{p}}))\mbox{ }\frac{1}{2} \left( 1 -
\lambda({\bf{p}}) \mbox{   }e^{ - 2 \sum_{ {\bf{q}} }\ll
A^{\dagger}_{ {\bf{p}}-{\bf{q}}/2 }({\bf{q}},t)A_{
{\bf{p}}-{\bf{q}}/2 }({\bf{q}},t) \gg } \right)e^{-i \epsilon_{ F
}(t-t^{'}) } E(|{\bf{p}}|-k_{F},t-t^{'}) \label{SURMISEG}
\end{equation}
\begin{equation}
<c^{\dagger}_{ {\bf{p}},< }(t^{'})c_{ {\bf{p}},< }(t)>  =
n_{F}({\bf{p}})\mbox{  }\frac{1}{2} \left( 1 + \lambda({\bf{p}})
\mbox{   }e^{ - 2 \sum_{ {\bf{q}} } \ll A^{\dagger}_{
{\bf{p}}+{\bf{q}}/2 }({\bf{q}},t)A_{ {\bf{p}}+{\bf{q}}/2
}({\bf{q}},t) \gg } \right)e^{-i \epsilon_{ F }(t-t^{'}) }
E(|{\bf{p}}|-k_{F},t^{'}-t) \label{SURMISEL}
\end{equation}
Here the envelope function has to be chosen with care. We make the
following definition which we justify a posteriori.
\begin{equation}
E(q,t-t^{'}) = \frac{ \int^{\infty}_{0} d \omega W(q,\omega) e^{
-i \omega(t-t^{'}) } }{ \int^{\infty}_{0} d \omega W(q,\omega) }
\end{equation}
where $ W(q,\omega) $ is the spectral weight,
\begin{equation}
W(q,\omega) = Im \left( \frac{1}{ \epsilon(q,\omega-i\delta) }
\right)
\end{equation}
In one dimension, the spectral weight is a delta function at the
collective mode. Hence in one dimension,
\begin{equation}
E(q,t-t^{'}) = e^{ -i \omega_{c}(q)(t-t^{'}) }
\end{equation}
In more than one dimension we have both the particle-hole mode and
collective mode. For example in the case of the jellium, the
collective mode contribution occurs at the plasma frequency and
this leads to a rapidly oscillating contribution which may be
ignored. The important contribution comes from the particle-hole
mode, which even though is not infinitely long-lived, makes a
significant contribution. For the noninteracting theory in any
number of dimensions this prescription gives us,
\begin{equation}
E_{free}(q,t-t^{'}) = e^{ -i v_{F}|q|(t-t^{'}) }
\end{equation}
 One may then use the Kubo-Martin-Schwinger (KMS)
boundary conditions to evaluate $ <c_{ {\bf{p}} }(t)c^{\dagger}_{
{\bf{p}} }(t^{'}) > $ from these propagators. First we observe
that this prescription reproduces the dynamical Green function of
the free theory in any number of dimensions both at finite
temperature and at zero temperature. We may use these to compute
the dynamical propagator of the Luttinger liquid and see if these
results agree with those of the more traditional approaches. Using
the traditional method,
\begin{equation}
<\psi_{R}^{\dagger}(x^{'},t^{'})\psi_{R}(0,t)> = e^{ - i
\epsilon_{F}(t-t^{'}) }\frac{ e^{ -i k_{F}x^{'} } }{ 2\pi i\mbox{
  }[x^{'} - v(t^{'}-t)] } \left[ \frac{ 1 }{ \Lambda |x^{'} -
v(t^{'}-t)| }\right]^{\gamma}
\end{equation}
Taking the Fourier transform with respect to $ x^{'} $ we obtain,
\begin{equation}
<c^{\dagger}_{p}(t^{'})c_{p}(t)> = e^{ -i \epsilon_{F} (t-t^{'})
}e^{ - i (|p|-k_{F})v(t-t^{'}) } \frac{1}{2}\left[ 1 + sgn( k_{F}
- |p|)\left(\frac{ |k_{F} - |p|| }{ \Lambda } \right)^{\gamma}
\right]
\end{equation}
This is completely identical to the result using Eq.(\ref{SURMISEG})
and Eq.(\ref{SURMISEL}) since $ \omega(q) = v|q| $. This approach
may seem quite ad-hoc but it would be very surprising indeed if a
theory that reproduces all dynamical aspects of the free theory both
at finite temperature and zero temperature correctly in any number
of dimensions and also able to reproduce the very nontrivial
dynamical propagator of the Luttinger liquid correctly in one
dimension is not valid in general. Besides, for the jellium, the
clever choice of the envelope function ensures that one-particle
Green function is gapless even though the collective mode is gapped.
Assuming that this is valid in general we may write down the
dynamical propagator of the Wigner crystal as follows.
\[
<c^{\dagger}_{p}(t^{'})c_{p}(t)>  = \frac{1}{2}\left( 1 +
\lambda(p) \mbox{   }e^{ - \frac{ m \omega_{0} }{k^2_{F}-p^2 }coth
\left( \frac{ \beta \omega_{0} }{2}\right) } \right)n_{F}(p) e^{-i
\epsilon_{ F }(t-t^{'}) }e^{ i \omega_{0}(t-t^{'}) }
\]
\begin{equation}
 + \frac{1}{2}\left( 1 -\lambda(p) \mbox{   } e^{ - \frac{ m \omega_{0}
}{p^2-k_{F}^2 }coth \left( \frac{ \beta \omega_{0} }{2}\right) }
\right)(1-n_{F}(p)) e^{-i \epsilon_{ F }(t-t^{'}) }e^{ - i
\omega_{0}(t-t^{'}) }
\end{equation}

\[
<c_{p}(t)c^{\dagger}_{p}(t^{'})>  = \frac{1}{2}\left( 1 -
\lambda(p) \mbox{   }e^{ - \frac{ m \omega_{0} }{k^2_{F}-p^2 }coth
\left( \frac{ \beta \omega_{0} }{2}\right) } \right)n_{F}(p) e^{-i
\epsilon_{ F }(t-t^{'}) }e^{ i \omega_{0}(t-t^{'}) }
\]
\begin{equation}
 + \frac{1}{2}\left( 1 +\lambda(p) \mbox{   } e^{ - \frac{ m \omega_{0}
}{p^2-k_{F}^2 }coth \left( \frac{ \beta \omega_{0} }{2}\right) }
\right)(1-n_{F}(p)) e^{-i \epsilon_{ F }(t-t^{'}) }e^{ - i
\omega_{0}(t-t^{'}) }
\end{equation}
Note that the above formulas are independent of arbitrarily chosen
momentum cutoffs and depend only on the microscopic parameters
present in the original hamiltonian with parabolic dispersion.
This in contrast with the conventional approaches to
bosonization\cite{HJSchulz} where such cutoffs are mandated by the
formalism. In this work, Schulz remarks that the time dependent
and temperature dependent formulas for the propagators are
complicated in his formalism and hence not too illuminating. This
is in contrast with our `momentum space bosonization' approach
where the above general formulas are not only simple but also
illuminating.

\section{ Conclusions}

To conclude, we have computed the dynamical Green function of the
Wigner crystal in one dimension whose momentum distribution
exhibits essential singularities in momentum space. We have
generalized the momentum distribution to finite temperature. We
have written down a general formula for the field operator without
any Klein factors in one dimension in terms of currents and
densities that does not involve momentum cutoffs and applies
directly to the Fermi gas with a parabolic dispersion rather than
to its caricature namely the Luttinger model. Lastly we have
summarized all the developments in the subject made by the author
and his collaborators in order to facilitate further developments.


\begin{thebibliography}{25}
\bibitem[16]{footnote} It is worth noting that issues such as `point splitting' \cite{review1} that
 occur in the traditional approaches to bosonization are due to peculiar nature
 of the Luttinger model that has right and left moving branches
 that extend all the way from minus infinity to plus infinity. We
 deal with the original hamiltonian with a parabolic dispersion
 and with a finite Fermi momentum and mass of electron, hence we do not have to address such issues.
\end{thebibliography}
\end{document}